\documentclass[epj]{svjour}
\pdfoutput=1

\usepackage{amsmath}
\usepackage{local}

\title{Topological phase transition in a network model with preferential attachment and node removal}
\author{
  Heiko Bauke\inst{1}
  \and Cristopher Moore\inst{2,3}\footnotemark[1]
  \and Jean-Baptiste Rouquier\inst{3,4}
  \and David Sherrington\inst{1,3}
}
\date{DOI: \href{http://dx.doi.org/10.1140/epjb/e2011-20346-0}{10.1140/epjb/e2011-20346-0}}

\institute{
  \href{http://www-thphys.physics.ox.ac.uk/}{Rudolf Peierls Centre for Theoretical Physics},
  \href{http://www.ox.ac.uk/}{University of Oxford},
  1 Keble Road, Oxford, OX1 3NP, United Kingdom
  \and
  \href{http://www.cs.unm.edu/}{Computer Science Department} and \href{http://panda.unm.edu/}{Department of Physics and Astronomy},
  \href{http://www.cs.unm.edu/}{University of New Mexico},
  Albuquerque, NM 87131, USA
  \and
  \href{http://www.santafe.edu}{Santa Fe Institute}, 1399 Hyde Park Road, Santa Fe, NM 87501, USA
  \and
  \href{http://www.ixxi.fr}{Rh{\^o}ne Alpes Complex Systems Institute},
  \href{http://www.ens-lyon.eu}{{\'E}cole Normale Sup{\'e}rieure de Lyon},
  \href{http://www.univ-lyon1.fr}{Universit{\'e} de Lyon 1},
  15 parvis Ren{\'e} Descartes,
  BP 7000
  69342 Lyon Cedex 07,
  France
}

\begin{document}

\abstract{ Preferential attachment is a popular model of growing
  networks.  We consider a generalized model with random node removal,
  and a combination of preferential and random attachment.  Using a
  high-degree expansion of the master equation, we identify a
  topological phase transition depending on the rate of node removal and
  the relative strength of preferential vs.\ random attachment, where
  the degree distribution goes from a power law to one with an
  exponential tail.
}

\maketitle
\renewcommand{\thefootnote}{\alph{footnote}}
\footnotetext[1]{\texttt{moore@cs.unm.edu}}

\section{Introduction}

Complex networks are found in nature, social and economic systems,
technical infrastructures, and countless other fields. The macroscopic
properties of such networks emerge from the microscopic interaction of
many individual constituents. Various models of complex networks have
been proposed and the statistical mechanics of networks has become an
established branch of statistical physics \cite{%
  Albert:Barabasi:2002:complex_networks,%
  Dorogovtsev:Mendes:2003:Evolution_of_networks,%
  Newman:2004:complex_networks,%
  Pastor-Satorras:Vespignani:2004:Evolution_and_Structure,%
  Boccaletti:etal:2006,Newman:Barabasi:Watts:2006:Networks}. Since
complex networks are non-equilibrium systems, they do not have to obey
detailed balance, and may show many fascinating features not found in
equilibrium systems.

In order to explain the power-law degree distribution observed in many
complex networks, Barab{\'a}si and
Albert~\cite{Barabasi:Albert:1999:Scaling_in_Random_Networks} introduced
a \emph{preferential attachment} model for growing networks.  When new
nodes enter the network, they prefer to attach to nodes with high
degree.  In a generalization of this model, the probability
$p_\mathrm{add}(v)$ that a new node $u$ establishes an edge to an
existing node $v$ with degree $k(v)$ is proportional to an
attractiveness function $A_k$; normalizing,
$p_\mathrm{add}(v)=A_{k(v)}/\sum_w A_{k(w)}$.  Barab{\'a}si and Albert
considered the case where the attachment is proportional to degree, $A_k
= k$.  This results in a power-law degree distribution $p_k \sim
k^{-\gamma}$ with exponent $\gamma=3$, which is close to the observed
exponents of many real world
networks~\cite{Newman:2004:complex_networks}.

However, the linear preferential attachment function of the
Barab{\'a}si-Albert model was introduced as an \emph{ad hoc} ansatz without
fundamental justification, and many generalizations are conceivable.
For many networks, a realistic model has to take into account node
removal, edge rewiring~\cite{Johnson:2009:rewiring} or
removal~\cite{Schneider:2011:depletion}, and other dynamical processes,
as well as deviations from linear preferential attachment.

Here we consider a generalized preferential attachment model with asymptotically
linear attractiveness function and random node removal.

The degree distribution of generalized preferential attachment models is
very sensitive to model-specific features.  Varying the
attractiveness
function~\cite{Dorogovtsev:Mendes:Samukhin:2000:Structure_of_Growing_Networks,%
  Krapivsky:Redner:2001:growing_networks,%
  Krapivsky:Rodgers:Redner:2001:Degree_Distributions,%
  Krapivsky:Redner:Leyvraz:2000:Growing_Networks,%
  Dorogovtsev:Mendes:2001:Scaling_properties}
or including node
removal~\cite{Dorogovtsev:Mendes:2000:Scaling_behaviour,%
  Sarshar:Roychowdhury:2004:complex_ad_hoc_networks,%
  Moore:Ghoshal:Newman:2006:Exact_solutions} can shift the exponent of
the power-law degree distribution to $2<\gamma<\infty$.  For networks of
constant size~\cite{Moore:Ghoshal:Newman:2006:Exact_solutions} or a
sublinear attractiveness
function~\cite{Krapivsky:Redner:2001:growing_networks}, the degree
distribution can become a stretched exponential.  We will demonstrate
that generalizations of the Barab{\'a}si-Albert model can dramatically
affect the degree distribution even in the case of growing networks and
asymptotically linear attractiveness, leading to a topological phase
transition from a power-law degree distribution to an exponential degree
distribution, with a stretched exponential at the critical point.

\section{The Model}

The topological phase transition in generalized preferential attachment
networks can be illustrated by considering the following model.
Vertices arrive at rate 1, each new vertex makes $c$ connections to
existing vertices, and we remove vertices randomly at a rate $r$.  Each
new edge attaches to a given pre-existing vertex of degree $k$ with
probability proportional to its attractiveness $A_k$.  We assume that
$A_k$ is of the form $$A_k=k+k^*$$ for some constant~$k^*$.  Thus the
choice of a link endpoint is somewhere between preferential attachment
(the case $k^*=0$) and uniform attachment (the limit $k^*\to\infty$).  
We can also treat $k^*$ as a the initial degree of the vertex when it is 
added to the network (e.g. by adding $k^*$ self-loops)  
and then run the ``pure'' preferential attachment model.

The network starts to grow at time $t=0$ with $N_0$ nodes and $M_0$
links.  We are interested in systems where the number of nodes is much
larger than unity.  Since we want to study the influence on the topology
of the dynamics of our model, and not the influence of the initial
network, we let the system evolve until we have added a number of nodes
much larger than $N_0$, and the degree distribution has reached
equilibrium.  For $r<1$ the number of nodes grows with time, so the
initial value $N_0$ is unimportant.  For $r=1$, nodes are added and
removed at the same rate; in that case the expected number of nodes is
constant, so we start with $N_0 \gg 1$.

\section{Analytic Solution}
\label{sec:analytic-solution}

In this section, we derive the average degree $\avk$ and the average
attractiveness $\avA$.  We then write the master equation for the degree
distribution $p_k$ and solve for its asymptotic behavior for large $k$
in terms of the parameters $r$, $c$, and $k^*$.

\subsection{Mean Degree and Attractiveness}

Let $p_k$ be the expected fraction of vertices in the network at a given
time that have degree~$k$.  As
in~\cite{Moore:Ghoshal:Newman:2006:Exact_solutions}, the expected mean
degree of a vertex $\avk=\sum_{k=0}^\infty kp_k$ can be derived as
follows.  The expected increase in the number of vertices per unit time
is $1-r$.  The expected number of edges removed when a randomly chosen
vertex is removed is $\avk$, so the expected increase in the number of
edges per unit time is $c-r\avk$.  At time $t$ the expected number of
vertices and edges are $n = (1 - r)t+N_0$ and
$m = (c - r\avk)t+M_0$, so in the limit $t \to \infty$ the mean degree obeys
\begin{equation*}
  \avk = \frac{2m}{n} = \frac{2(c-r\avk)}{1-r} \, ,
\end{equation*}
and solving for $\avk$ gives 
\begin{equation}
  \label{eq:avk}
  \avk = \frac{2c}{1+r} \, .
\end{equation}
The average attractiveness is then 
\begin{equation}
  \label{eq:avA}
  \avA = \sum_{k=0}^\infty A_k p_k = \avk + k^* = \frac{2c}{1+r} + k^* \, .
\end{equation}
In the case $r=1$ where the network has constant size, we have $\avk=c$
and $\avA = c+k^*$.

\subsection{Master Equation}

Let $n$ be the number of vertices at time $t$.  The expected number of
vertices with degree $k$ is $n_k = np_k$.  One time step later this is
$n'_k = (n+1-r)p'_k$ where $p'_k$ is the new value of $p_k$. Thus
\begin{multline}
  (n + 1 - r)p'_k 
  = np_k + \delta_{kc} 
  + \frac{c}{\avA} \left( A_{k-1} p_{k-1} - A_k p_k \right) \\
  + r (k+1) p_{k+1} - r k p_k - rp_k \, . \label{eq:master:growing}
\end{multline}
The term $\delta_{kc}$ corresponds to adding of a vertex of degree $c$
to the network.  The term $c A_{k-1}p_{k-1} / \avA$ is the probability
that a vertex of degree $k-1$ gains an extra edge from the new vertex
and becomes of degree $k$, and similarly $c A_{k-1}p_{k-1} / \avA$ is
the flow from degree $k$ to degree $k+1$.  The terms $r(k+1)p_{k+1}$ and
$rkp_k$ are the flows from $k+1$ to $k$ and from $k$ to $k-1$
respectively, as vertices lose edges when one of their neighbors is
removed from the network.  Finally, $rp_k$ is the probability that a
vertex of degree $k$ is removed.  Contributions from processes in which
a vertex gains or loses two or more edges in a single unit of time
vanish in the limit of large $n$ and have been neglected.

We are interested in the asymptotic form of the degree distribution
$p_k$ in the limit of large $t$.  Setting $p'_k = p_k$
in~\eqref{eq:master:growing} gives
\begin{multline}
  \delta_{k,c} + \frac{c}{\avA} \left( A_{k-1} p_{k-1} - A_k p_k \right) \\
  + r (k+1) p_{k+1} - r k p_k - p_k = 0 \, . \label{eq:master}
\end{multline}
as previously appeared in~\cite{Moore:Ghoshal:Newman:2006:Exact_solutions}.  Naively, this equation appears linear in the $p_k$.  But since it involves the mean attractiveness $\avA$, the combination of~\eqref{eq:avA} and~\eqref{eq:master} gives a nonlinear system of equations.

When the attractiveness is proportional to the degree, $A_k=k$, the
system~\eqref{eq:avA} and \eqref{eq:master} separates, and has been
solved analytically in~\cite{Moore:Ghoshal:Newman:2006:Exact_solutions}.
The authors showed that in this case the degree distribution exhibits a
power-law tail in the case $0 \le r < 1$ of growing networks, and
follows a stretched exponential in the constant-size case $r=1$.  For
more general attractiveness functions like the one in this paper, a
fully analytic solution seems more difficult.  Thus we focus on the
behavior of $p_k$ for large $k$.
Depending on the model parameters we find either a degree distribution
with a power-law tail (Figure~\ref{fig/compareB}) or an exponential tail
(Figure~\ref{fig/compareA3}).  We confirm our calculations with
numerical simulation of the master equation, and through direct
simulation of the network dynamics.

\subsection{High-Degree Expansion}
\label{sec:solution}

We now specialize the master equation to our model.
Substituting~\eqref{eq:avA} into~\eqref{eq:master} gives, for $k \notin
\{c-1,c,c+1\}$,
\begin{multline}
  \label{eq:master-model}
  \frac{c}{\avA} \big( (k-1+k^*) p_{k-1} - (k+k^*) p_k \big) \\
  + r (k+1) p_{k+1} - r k p_k - p_k = 0 \, .
\end{multline}
We will determine the asymptotic behavior of $p_k$ with a ``high-degree
expansion'', by approximating the ratio $p_k / p_{k-1}$ as a Taylor
series in $1/k$.  We find a phase transition between power-law and
exponential behavior, and determine the phase diagram as a function of
the parameters $c$, $r$, and $k^*$.

As an ansatz, assume that $p_k$ is a power-law times an exponential:
\begin{equation}
  \label{eq:ansatz}
  p_k = C k^\alpha \beta^k \, .
\end{equation}
We can determine $\alpha$ and $\beta$ by taking $k \gg 1$, and expanding
the ratio $p_k / p_{k-1}$ to leading orders in $1/k$.  This gives
\begin{align}
  \label{eq:pk_pk1}
  \frac{p_k}{p_{k-1}} & = 
  \beta \left(
    1+\frac{\alpha}{k} 
    + O\!\left(\frac1{k^2}\right)
  \right) \, , \\
  \frac{p_{k+1}}{p_{k-1}} & = 
  \beta^2 \left(
    1+\frac{2\alpha}{k} 
    + O\!\left(\frac1{k^2}\right)
  \right) \, .
\end{align}
Substituting this into~\eqref{eq:master-model}, multiplying by
$\avA/p_{k-1}$, and ignoring $O(1/k)$ terms
yields the equation
\begin{multline}
  k  (\beta-1)(r \avA \beta - c) 
  + \alpha \big(  r \avA \beta(2\beta-1)  -c\beta\big) \\
  + \avA \beta (r \beta-1) + c \big(k^*(1-\beta)-1\big) = 0 \, . \label{eq:collect}
\end{multline}
Since~\eqref{eq:collect} must be true for all $k$, we can set the
coefficient of $k$ to zero.  This gives two solutions for $\beta$,
namely $\beta=1$ and
\begin{equation}
  \label{eq:b}
  \beta = \frac{c}{r \avA} \, .
\end{equation}

\begin{figure}
  \centering
  \resizebox{\columnwidth}{!}{\large
\begingroup
  \makeatletter
  \providecommand\color[2][]{%
    \GenericError{(gnuplot) \space\space\space\@spaces}{%
      Package color not loaded in conjunction with
      terminal option `colourtext'%
    }{See the gnuplot documentation for explanation.%
    }{Either use 'blacktext' in gnuplot or load the package
      color.sty in LaTeX.}%
    \renewcommand\color[2][]{}%
  }%
  \providecommand\includegraphics[2][]{%
    \GenericError{(gnuplot) \space\space\space\@spaces}{%
      Package graphicx or graphics not loaded%
    }{See the gnuplot documentation for explanation.%
    }{The gnuplot epslatex terminal needs graphicx.sty or graphics.sty.}%
    \renewcommand\includegraphics[2][]{}%
  }%
  \providecommand\rotatebox[2]{#2}%
  \@ifundefined{ifGPcolor}{%
    \newif\ifGPcolor
    \GPcolortrue
  }{}%
  \@ifundefined{ifGPblacktext}{%
    \newif\ifGPblacktext
    \GPblacktexttrue
  }{}%
  \let\gplgaddtomacro\g@addto@macro
  \gdef\gplbacktext{}%
  \gdef\gplfronttext{}%
  \makeatother
  \ifGPblacktext
    \def\colorrgb#1{}%
    \def\colorgray#1{}%
  \else
    \ifGPcolor
      \def\colorrgb#1{\color[rgb]{#1}}%
      \def\colorgray#1{\color[gray]{#1}}%
      \expandafter\def\csname LTw\endcsname{\color{white}}%
      \expandafter\def\csname LTb\endcsname{\color{black}}%
      \expandafter\def\csname LTa\endcsname{\color{black}}%
      \expandafter\def\csname LT0\endcsname{\color[rgb]{1,0,0}}%
      \expandafter\def\csname LT1\endcsname{\color[rgb]{0,1,0}}%
      \expandafter\def\csname LT2\endcsname{\color[rgb]{0,0,1}}%
      \expandafter\def\csname LT3\endcsname{\color[rgb]{1,0,1}}%
      \expandafter\def\csname LT4\endcsname{\color[rgb]{0,1,1}}%
      \expandafter\def\csname LT5\endcsname{\color[rgb]{1,1,0}}%
      \expandafter\def\csname LT6\endcsname{\color[rgb]{0,0,0}}%
      \expandafter\def\csname LT7\endcsname{\color[rgb]{1,0.3,0}}%
      \expandafter\def\csname LT8\endcsname{\color[rgb]{0.5,0.5,0.5}}%
    \else
      \def\colorrgb#1{\color{black}}%
      \def\colorgray#1{\color[gray]{#1}}%
      \expandafter\def\csname LTw\endcsname{\color{white}}%
      \expandafter\def\csname LTb\endcsname{\color{black}}%
      \expandafter\def\csname LTa\endcsname{\color{black}}%
      \expandafter\def\csname LT0\endcsname{\color{black}}%
      \expandafter\def\csname LT1\endcsname{\color{black}}%
      \expandafter\def\csname LT2\endcsname{\color{black}}%
      \expandafter\def\csname LT3\endcsname{\color{black}}%
      \expandafter\def\csname LT4\endcsname{\color{black}}%
      \expandafter\def\csname LT5\endcsname{\color{black}}%
      \expandafter\def\csname LT6\endcsname{\color{black}}%
      \expandafter\def\csname LT7\endcsname{\color{black}}%
      \expandafter\def\csname LT8\endcsname{\color{black}}%
    \fi
  \fi
  \setlength{\unitlength}{0.0500bp}%
  \begin{picture}(7200.00,5040.00)%
    \gplgaddtomacro\gplbacktext{%
      \csname LTb\endcsname%
      \put(1210,704){\makebox(0,0)[r]{\strut{} 0}}%
      \put(1210,1518){\makebox(0,0)[r]{\strut{} 20}}%
      \put(1210,2333){\makebox(0,0)[r]{\strut{} 40}}%
      \put(1210,3147){\makebox(0,0)[r]{\strut{} 60}}%
      \put(1210,3962){\makebox(0,0)[r]{\strut{} 80}}%
      \put(1210,4776){\makebox(0,0)[r]{\strut{} 100}}%
      \put(1342,484){\makebox(0,0){\strut{} 0}}%
      \put(2448,484){\makebox(0,0){\strut{} 0.2}}%
      \put(3553,484){\makebox(0,0){\strut{} 0.4}}%
      \put(4659,484){\makebox(0,0){\strut{} 0.6}}%
      \put(5764,484){\makebox(0,0){\strut{} 0.8}}%
      \put(6870,484){\makebox(0,0){\strut{} 1}}%
      \put(440,2740){\rotatebox{90}{\makebox(0,0){\strut{}$k^*$}}}%
      \put(4106,154){\makebox(0,0){\strut{}$r$}}%
    }%
    \gplgaddtomacro\gplfronttext{%
      \csname LTb\endcsname%
      \put(1342,1600){\makebox(0,0)[l]{\strut{}\begin{tabular}{l}power law\\ degree distribution\end{tabular}}}%
      \put(4659,3147){\makebox(0,0)[l]{\strut{}\begin{tabular}{c}exponential\\degree distribution\end{tabular}}}%
    }%
    \gplbacktext
    \put(0,0){\includegraphics{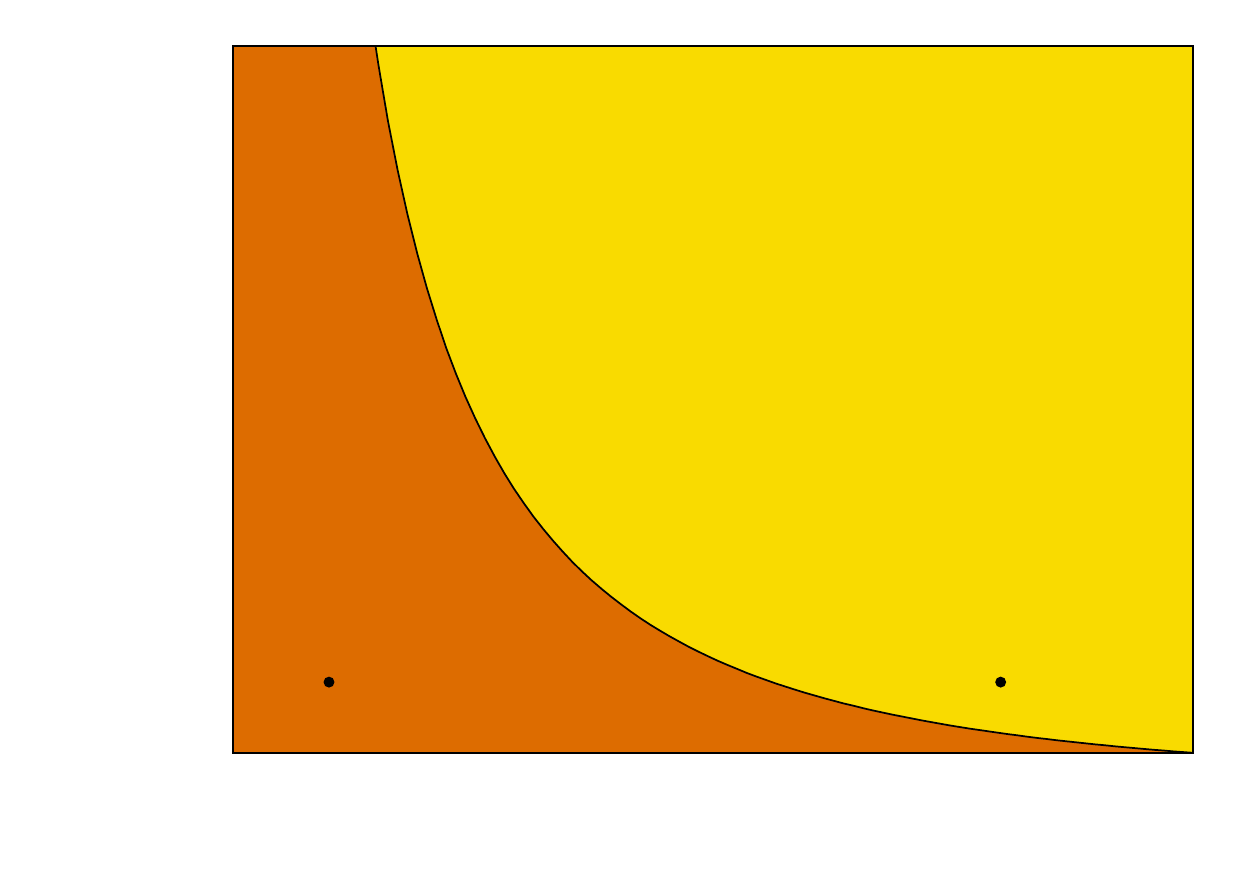}}%
    \gplfronttext
  \end{picture}%
\endgroup
}
  \caption{Phase diagram of the degree distribution with $c = 20$.  The
    two black dots indicate locations in the phase space for data shown
    on Figures~\ref{fig/compareB}, \ref{fig/compareA3}
    and~\ref{fig/compareA2}.}
  \label{fig/phase-diag}
\end{figure}

If $r \avA > c$, the solution $\beta < 1$ of~\eqref{eq:b} is physically
relevant and $p_k$ decays exponentially.  However, if \mbox{$r \avA <
  c$} then~\eqref{eq:b} would give $\beta > 1$, which does not
correspond to a normalizable probability distribution.  In that case
$\beta=1$ is the relevant solution, and $p_k \sim k^\alpha$
is a power-law.  Thus a phase transition occurs at $r \avA = c$.
Applying~\eqref{eq:avA}, we can write this in terms of a critical value
of $k^*$,
\begin{equation}
  k^*_c = \frac{c(1-r)}{r(1+r)} \, .
\end{equation}
We illustrate the resulting phase diagram in Figure~\ref{fig/phase-diag}.

\subsection{The Power-Law}
\label{sec:below}

To solve for the power-law exponent $\alpha$, we again
use~\eqref{eq:collect}, but now set the constant term (with respect to
$k$) to zero.  If $\beta=1$, this gives $p_k \sim k^\alpha$ where
\begin{equation}
  \label{eq:alpha1}
  \alpha = -\frac{\avA(1-r)+c}{c-\avA r} < 0 \, .
\end{equation}
Note that $\alpha$ approaches $-\infty$ as we approach the transition.
As we show in Section~\ref{sec:stretched}, at criticality $p_k$ takes a
stretched-exponential form.

Substituting~\eqref{eq:avk} and~\eqref{eq:avA} into~\eqref{eq:alpha1},
we can express $\alpha$ in terms of $c$, $r$, and $k^*$ as
\begin{equation*}
  \alpha=-\frac{c(3-r)+k^* (1-r^2)}{c(1-r)-k^* r(1+r)}\,.
\end{equation*}
In the special case $k^*=0$, this recovers the result
of~\cite{Moore:Ghoshal:Newman:2006:Exact_solutions}
\begin{equation*}
  \alpha = -\frac{3-r}{1-r}\,,
\end{equation*}
while in the special case $r=0$, it includes the result
of~\cite{Dorogovtsev:Mendes:Samukhin:2000:Structure_of_Growing_Networks}
\begin{equation*}
  \alpha = -3-\frac{k^*}{c}\,.
\end{equation*}
Finally, since $k^*\geq 0$ and $r\in[0,1]$, we have 
\begin{equation*}
  -\infty < \alpha \le -3 \, ,
\end{equation*}
so that the degree distribution has a finite average as well as (except
when $k^*=r=0$) a finite variance.

Above the transition, $\beta$ is given by~\eqref{eq:b}.  Again setting
the constant term of~\eqref{eq:collect} to zero gives
\begin{equation}
  \label{eq:alphab}
  \alpha = -\frac{c(k^*-1)+\avA(1+r(1-k^*))}{\avA r-c} \, . 
\end{equation}
Thus we have a power-law correction to the exponential decay, $p_k \sim
k^\alpha \beta^k$.  In terms of our parameters,
\begin{gather*}
  \label{eq:above-1st-order}
  \beta = \frac{c(1+r)}{r(2c + k^* (1+r))} \\
  \alpha = k^* - \frac{(1+r)(c+k^* (1+r))}{k^* r(1+r)-c(1-r)} \, . 
\end{gather*}
Note that in this regime $\alpha$ may be positive.

\subsection{Finite-Degree Corrections}
\label{sec:above-transition}

We have derived the leading behavior of $p_k$ for large $k$, namely a
power-law times an exponential.  In this section, we obtain the
next-order correction, by taking the Taylor series to second order in
$1/k$.  This correction becomes important in the exponential regime, where
the exponential decay of the degree distribution makes smaller degrees
more relevant.

Rather than starting with an ansatz for the correction term, we derive it by 
expanding $p_k/p_{k-1}$ to second order in $1/k$.  Write
\begin{equation}
  \label{eq:pk-2nd-order}
  \frac{p_k}{p_{k-1}} = \beta \left( 
    1+ \frac{\alpha}{k} + \frac{\kappa}{k^2} 
    + O\!\left(\frac1{k^3}\right) \!
  \right) \, .
\end{equation}
For $k > c$, we then have
\begin{align}
  p_k
  &\sim \beta^k \prod_{i=1}^k \left( 
    1+ \frac{\alpha}{i} + \frac{\kappa}{i^2} + O\!\left(\frac1{i^3}\right) \!
  \right) \nonumber \\
  &= \beta^k \exp\!\left( \sum_{i=1}^k \ln \!\left( 
      1+ \frac{\alpha}{i} + \frac{\kappa}{i^2} + O\!\left(\frac1{i^3}\right)\!
    \right) \!
  \right) \nonumber \\
  &= \beta^k \exp\!\left( \sum_{i=1}^k \left( 
      \frac{\alpha}{i} + \frac{\kappa-\alpha^2/2}{i^2} + O\!\left(\frac1{i^3}\right)\! \right) \!
  \right) \nonumber \\
  &\sim \beta^k \exp\!\left( 
    \alpha \ln k + \frac{(\alpha+\alpha^2)/2 - \kappa}{k} + O\!\left(\frac1{k^2}\right)\!
  \right) \, , 
  \label{eq:kappa}
\end{align}
where $\sim$ hides multiplicative constants.  
Here we used the Taylor series for the logarithm, 
\begin{equation*} 
    \ln (1+\eps) = \eps - \eps^2/2 + O(\eps^3) \, , 
\end{equation*}
the approximation for the $k$th harmonic number
\begin{equation*}
  \sum_{i=1}^k \frac{1}{i} = \ln k + \gamma + \frac{1}{2k} + O(1/k^2) \,,
\end{equation*}
where $\gamma$ is Euler's constant, and
\begin{equation*}
  \sum_{i=1}^k \frac{1}{i^2} = \frac{\pi^2}{6} - \frac{1}{k} + O(1/k^2) \, .
\end{equation*}

Eq.~\eqref{eq:kappa} gives a multiplicative correction of the form $\e^{\delta/k}$ to our earlier
form~\eqref{eq:ansatz} for $p_k$,
\begin{equation}
  p_k = C \beta^k k^\alpha \e^{\delta/k} \left( 1+O(1/k^{2}) \right) \, , 
  \label{eq:pk-as-f-of-k}
\end{equation}
where
\begin{equation}
\label{eq:kappa-delta}
\delta = \frac{\alpha+\alpha^2}{2} - \kappa \, . 
\end{equation}
To determine $\kappa$ and therefore $\delta$, using~\eqref{eq:pk-2nd-order} we write
\begin{equation}
  \label{eq:p1_p_1-2nd-order}
  \frac{p_{k+1}}{p_{k-1}} = \beta^2  \left(1 + \frac{2\alpha}{k} + \frac{2\kappa + \alpha^2}{k^2} + O\!\left(\frac{1}{k^3}\right)\!\right) \, .   
\end{equation}
Substituting \eqref{eq:pk-2nd-order} and~\eqref{eq:p1_p_1-2nd-order} 
into the master equation~\eqref{eq:master-model} and multiplying by
$\avA / p_{k-1}$,
we obtain an equation akin to~\eqref{eq:collect}, with terms of order
$k$, $1$, $1/k$, and negligible terms of order $O(1/k^2)$.  The
coefficient of $k$ and the constant term are identical to those
in~\eqref{eq:collect}, giving the same solutions for $\alpha$ and
$\beta$ as before.  Setting the coefficient of $1/k$ to zero 
and applying~\eqref{eq:kappa-delta} gives
\begin{equation*}
  \delta = \frac{\alpha}{2} \frac{c(1+\alpha+k^*) + \avA(2 + r(1 + \alpha - 6 \beta - 4 \alpha \beta))}{c + r \avA (1-2\beta)} \, . 
\end{equation*}

Multiplicatively speaking, the correction term $\e^{\delta/k}$ becomes
negligible as $k \to \infty$.  However, it makes a significant
difference for small values of $k$, and greatly improves agreement with
the simulations in the next section.  We note that the same technique,
expanding the ratio between $p_{k-1}$, $p_k$, and $p_{k+1}$ to higher
degree in $1/k$, can give us as many correction terms as we wish.

\subsection{The Stretched Exponential At Criticality}
\label{sec:stretched}

We saw above that as we approach the critical point, the power-law
exponent $\alpha$ diverges to $-\infty$, and the exponential factor
$\beta$ approaches $1$.  In this section we show that the degree
distribution in fact becomes a stretched exponential at this point, due
to the appearance of half-integer powers of $1/k$ in the high-degree
expansion of $p_k/p_{k-1}$.

We start with the ansatz 
\begin{equation*}
  p_k = C k^\alpha \beta^k \,\e^{\zeta \sqrt{k}} \, .
\end{equation*}
Expanding to order $k^{-3/2}$, we have
\begin{align*}
  \frac{p_k}{p_{k-1}} 
  &= \beta \left( 1 + \frac{\zeta}{2 \sqrt{k}} + \frac{\alpha+\zeta^2/8}{k} + \frac{\zeta^3/48 + \alpha \zeta/2 + \zeta/8}{k^{3/2}} \right) \\
  \frac{p_{k+1}}{p_{k-1}} 
  &= \beta^2 \left( 1 + \frac{\zeta}{\sqrt{k}} + \frac{2 \alpha+\zeta^2/2}{k} + \frac{\zeta^3/6 + 2 \alpha \zeta}{k^{3/2}} \right) \, ,
\end{align*}
with error terms of order $1/k^2$.  Substituting this into the master
equation~\eqref{eq:master-model} as before, if $\zeta \ne 0$ then the
terms of order $k$ and $\sqrt{k}$ force $b=1$ and $c=r \avA$.  In other
words, $\zeta$ can be nonzero only at the critical point.  The term of
order $1$ then gives
\begin{equation*}
  \zeta = -2/\sqrt{r} \, , 
\end{equation*}
and the term of order $k^{-1/2}$ gives the power-law correction
\begin{equation*}
  \alpha = -\frac{3}{4} + \frac{k^*}{2} \, . 
\end{equation*}

This recovers the results
of~\cite{Moore:Ghoshal:Newman:2006:Exact_solutions} for the special case
$k^*=0$ and $r=1$, where $\zeta = -2$ and $\alpha = -3/4$.  However,
these calculations are significantly more technical, evaluating
generating functions and their derivatives in terms of special
functions.  Our high-degree expansion is closer in spirit
to~\cite{Krapivsky:Redner:Leyvraz:2000:Growing_Networks}, where $p_k$ is
written as a telescoping product of ratios $p_k/p_{k-1}$.

\section{Simulations}
\label{sec:simulations}

To check that our asymptotic calculations are correct, we conducted two
kinds of simulations: direct simulation of the dynamics of finite
networks, and numerical integration of the master equation.

In our direct simulations, we grew the network stochastically according
to the model, up to size $n = 5 \times 10^7$.  However, it is hard to
explore the tail of the degree distribution in the exponential regime,
since $p_k$ falls off exponentially.  For instance, to measure a
probability $p_k \approx 10^{-10}$ we would need a network of size more
than $n = 10^{10}$, unless we use large bin sizes.  In this case,
numerically integrating the master equation~\eqref{eq:master:growing}
until it reaches equilibrium
lets us explore the asymptotics of $p_k$ far more efficiently.

Since the normalization constant $C$ depends on the values of $p_k$ for
small $k$, which our analysis does not try to predict, we adjust $C$ to
fit the simulations.  We do not tune any other parameters.  In
particular, $\alpha$ and $\beta$ are determined by our analysis, rather
than fit to the data.

\begin{figure}
  \centering\noindent
  \resizebox{\columnwidth}{!}{\hspace{-0.8cm}\large\input{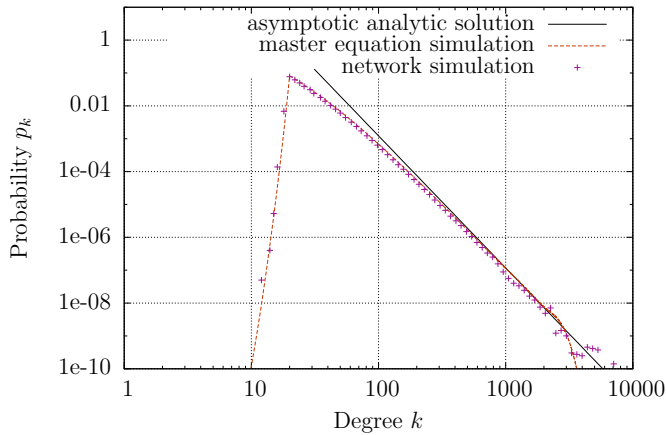}}
  \caption{Comparison between simulations for $k^*=10$, $r=0.10$, $c=20$
    and our solution, which gives a power-law $p_k \sim k^\alpha$ with
    $\alpha = -4.02$.  There is good agreement for $k>200$.  This set of
    parameters is in the power-law regime, below the phase transition at
    $k^*_c \approx 163.6$.  Note the log-log scale.  The integrated
    master equation dips down at $k \approx 3000$ because it has not yet
    reached equilibrium at the highest degrees. \newline We show a
    network simulation for a single network.  For large degrees, we use
    logarithmic binning, so that each point represents the average over
    an interval of degrees of width proportional to $\log k$.  }
  \label{fig/compareB}
\end{figure}

Figure~\ref{fig/compareB} shows results in the power-law regime below
the transition.  There is good agreement between our solution and both
types of simulations above $k > 200$ or so.  The direct simulation
differs somewhat from the master equation at large $k$ due to
finite-size effects.

\begin{figure}
  \centering
  \resizebox{\columnwidth}{!}{\hspace{-0.8cm}\large\input{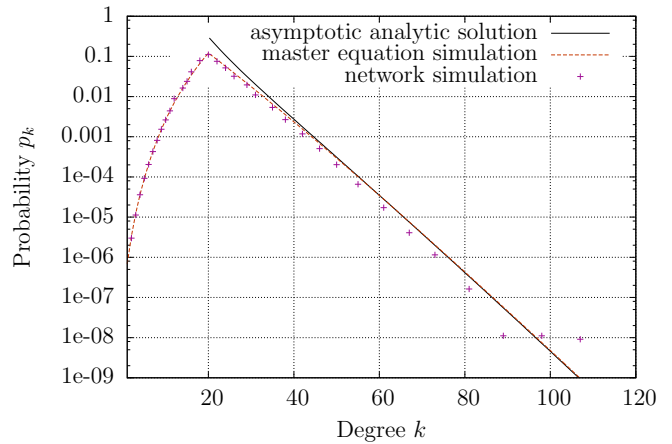}}
  \caption{Comparison between simulations for $k^*=10$, $r=0.80$, $c=20$
    and our solution, which gives $p_k \sim \beta^k k^\alpha
    \mathrm{e}^{\delta/k}$ with $\beta = 0.776$, $\alpha = 3.423$ and
    $\delta=80.89$.  There is good agreement for $k>30$.  This set of
    parameters is in the exponential regime, above the phase transition
    at $k^*_c \approx 2.78$.  Note the semi-logarithmic scale.  }
  \label{fig/compareA3}
\end{figure}

\begin{figure}
  \centering
  \resizebox{\columnwidth}{!}{\large\input{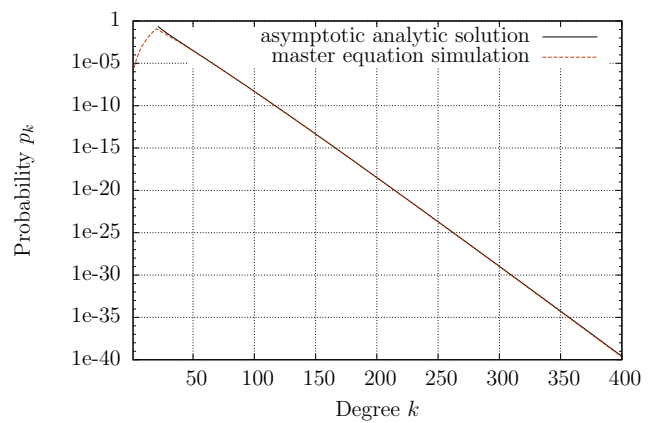}}
  \caption{Comparison of our analytic solution and the integrated master
    equation for $k^*=10$, $r=0.80$, $c=20$ (the same parameters as
    Figure~\ref{fig/compareA3}) at larger degrees than direct
    simulations can reach.}
  \label{fig/compareA2}
\end{figure}

Figure~\ref{fig/compareA3} shows results in the exponential regime,
above the phase transition.  Here as well, there is good agreement
between simulations and our asymptotic solution for large enough $k$.
Figure~\ref{fig/compareA2} shows the same parameters at larger degrees;
as discussed above, we reach these larger degrees by abandoning direct
simulation and integrating the master equation.  The agreement with our
asymptotic solution is 
excellent.

\begin{sloppypar}
  The pseudo-code and source code for the simulations can be found at
  \url{http://www.rouquier.org/jb/research/papers/2010_growing_network/}.
\end{sloppypar}

\section{Conclusion}

We have studied dynamical networks that are generated by a model 
where growth takes place through a combination of 
preferential and uniform attachment, and where nodes are removed randomly
at a certain rate. 
Both growth and node removal are key features of many real-world networks.  
Nodes in peer-to-peer networks may be added or removed, 
people join and leave social networks, nodes in a communication network
can be attacked or degrade with time, and so on.  Uniformly random
attachment appears, for instance, if people choose random seats and 
strike up conversations with their neighbors, or are assigned to random classrooms; 
random attachment also occurs, by design, in some peer-to-peer protocols.

We have solved for the asymptotic degree distribution, and found a phase
transition between power-law and exponential behavior, with a stretched exponential 
at the critical point.  Thus, in
contrast to pure growth models, an asymptotically linear attractiveness
function is not a sufficient condition for a power-law degree
distribution. If the growth rate is too small, or the node removal rate is too
high, the degree distribution is exponential.

Our findings are relevant for real-world networks where both growth and node removal are important.  They also imply further potentially interesting consequences for the evolution of networks whose effective growth rates vary with time; for instance, for networks that start with a high growth rate, but that reach a state where nodes are added and removed at about the same rate, e.g. due to limits on the network's overall size or population.

Similarly, it would be interesting to understand how the macro-dynamics of a network change as the growth and/or removal rates are varied so that we approach or cross the topological phase transition.  This includes the approach to the asymptotic degree distribution from a nonequilibrium initial state; whether this approach shows critical slowing down near the transition; and the dynamics after a sudden change of the growth and/or removal rates, say from a region from the power-law regime to the exponential one.

A broader question is how the transition
affects various types of dynamics taking place on the network, such as
search~\cite{adamic2001search}, congestion, and robustness to attack.

Finally, another direction for future work is to introduce some kind of quenched disorder into the network model.  
This could include allowing $k^*$ to vary from node to node, based on the node's 
intrinsic ``fitness'' or ``attractiveness''~\cite{Dorogovtsev:Mendes:2001:Scaling_properties,Bianconi:Barabasi:2001:Competition_and_multiscaling}.  
We believe that the asymptotic behavior of the degree distribution and its phase diagram will be similar to our results here as long as $k^*$ has bounded expectation and variance.

\subsection*{Acknowledgments}
This work has been partly sponsored by the European Community's FP6
Information Society Technologies programme under contract IST-001935,
EVERGROW.  C. M. is supported by the McDonnell Foundation.
D. S. acknowledges support from the Leverhulme Trust in the form of an Emeritus Fellowship.
We are also
grateful to the Santa Fe Institute who hosted the authors and fostered
our collaboration.

\bibliographystyle{epj_doi}
\bibliography{biblio-doi}

\begin{thebibliography}{19}
\providecommand{\doi}[1]{\href{http://dx.doi.org/#1}{DOI: \nolinkurl{#1}}}
\providecommand{\nolinkurl}[1]{\texttt{#1}}
\providecommand{\href}[2]{#2: \url{#1}}
\providecommand{\url}[1]{\texttt{#1}}

\bibitem{Albert:Barabasi:2002:complex_networks}
R.~Albert, A.L. Barab\'asi.
\newblock \emph{Statistical mechanics of complex networks}.
\newblock Reviews of Modern Physics \textbf{74}(1), 47 (2002).
\newblock \doi{10.1103/RevModPhys.74.47}

\bibitem{Dorogovtsev:Mendes:2003:Evolution_of_networks}
S.N. Dorogovtsev, J.F.F. Mendes, \emph{Evolution of Networks From Biological
  Nets to the Internet and WWW} (Oxford University Press, 2003)

\bibitem{Newman:2004:complex_networks}
M.E.J. Newman.
\newblock \emph{The structure and function of complex networks}.
\newblock SIAM Review \textbf{45}(2), 167 (2003).
\newblock \doi{10.1137/S003614450342480}

\bibitem{Pastor-Satorras:Vespignani:2004:Evolution_and_Structure}
R.~Pastor-Satorras, A.~Vespignani, \emph{Evolution and Structure of the
  Internet: A Statistical Physics Approach} (Cambridge University Press,
  Cambridge, 2004)

\bibitem{Boccaletti:etal:2006}
S.~Boccaletti, V.~Latora, Y.~Moreno, M.~Chavez, D.U. Hwang.
\newblock \emph{Complex networks: Structure and dynamics}.
\newblock Physics Reports \textbf{424}(4--5), 175 (2006).
\newblock \doi{10.1016/j.physrep.2005.10.009}

\bibitem{Newman:Barabasi:Watts:2006:Networks}
M.E.J. Newman, A.L. Barab\'asi, D.J. Watts, \emph{The Structure and Dynamics of
  Networks} (Princeton University Press, Princeton, 2006)

\bibitem{Barabasi:Albert:1999:Scaling_in_Random_Networks}
A.L. Barab\'asi, R.~Albert.
\newblock \emph{Emergence of scaling in random networks}.
\newblock Science \textbf{286}(5439), 509 (1999).
\newblock \doi{10.1126/science.286.5439.509}

\bibitem{Johnson:2009:rewiring}
S.~Johnson, J.~Torres, J.~Marro.
\newblock \emph{Nonlinear preferential rewiring in fixed-size networks as a
  diffusion process}.
\newblock Physical Review E \textbf{79}(5), 050104 (2009).
\newblock \doi{10.1103/PhysRevE.79.050104}

\bibitem{Schneider:2011:depletion}
C.~Schneider, L.~de~Arcangelis, H.~Herrmann.
\newblock \emph{Scale-free networks by preferential depletion}.
\newblock EPL (Europhysics Letters) \textbf{95}, 16005 (2011).
\newblock \doi{10.1209/0295-5075/95/16005}

\bibitem{Dorogovtsev:Mendes:Samukhin:2000:Structure_of_Growing_Networks}
S.N. Dorogovtsev, J.F.F. Mendes, A.N. Samukhin.
\newblock \emph{Structure of growing networks with preferential linking}.
\newblock Physical Review Letters \textbf{85}(21), 4633 (2000).
\newblock \doi{10.1103/PhysRevLett.85.4633}

\bibitem{Krapivsky:Redner:2001:growing_networks}
P.L. Krapivsky, S.~Redner.
\newblock \emph{Organization of growing random networks}.
\newblock Physical Review {E} \textbf{63}(6), 066123 (2001).
\newblock \doi{10.1103/PhysRevE.63.066123}

\bibitem{Krapivsky:Rodgers:Redner:2001:Degree_Distributions}
P.L. Krapivsky, G.J. Rodgers, S.~Redner.
\newblock \emph{Degree distributions of growing networks}.
\newblock Physical Review Letters \textbf{86}(23), 5401 (2001).
\newblock \doi{10.1103/PhysRevLett.86.5401}

\bibitem{Krapivsky:Redner:Leyvraz:2000:Growing_Networks}
P.L. Krapivsky, S.~Redner, F.~Leyvraz.
\newblock \emph{Connectivity of growing random networks}.
\newblock Physical Review Letters \textbf{85}(21), 4629 (2000).
\newblock \doi{10.1103/PhysRevLett.85.4629}

\bibitem{Dorogovtsev:Mendes:2001:Scaling_properties}
S.N. Dorogovtsev, J.F.F. Mendes.
\newblock \emph{Scaling properties of scale-free evolving networks: Continuous
  approach}.
\newblock Physical Review {E} \textbf{63}(5), 056125 (2001).
\newblock \doi{10.1103/PhysRevE.63.056125}

\bibitem{Dorogovtsev:Mendes:2000:Scaling_behaviour}
S.N. Dorogovtsev, J.F.F. Mendes.
\newblock \emph{Scaling behaviour of developing and decaying networks}.
\newblock Europhysics Letters \textbf{52}(1), 33 (2000).
\newblock \doi{10.1209/epl/i2000-00400-0}

\bibitem{Sarshar:Roychowdhury:2004:complex_ad_hoc_networks}
N.~Sarshar, V.~Roychowdhury.
\newblock \emph{Scale-free and stable structures in complex ad hoc networks}.
\newblock Physical Review {E} \textbf{69}(2), 026101 (2004).
\newblock \doi{10.1103/PhysRevE.69.026101}

\bibitem{Moore:Ghoshal:Newman:2006:Exact_solutions}
C.~Moore, G.~Ghoshal, M.E.J. Newman.
\newblock \emph{Exact solutions for models of evolving networks with addition
  and deletion of nodes}.
\newblock Physical Review~E \textbf{74}(3), 036121 (2006).
\newblock \doi{10.1103/PhysRevE.74.036121}

\bibitem{adamic2001search}
L.~Adamic, R.~Lukose, A.~Puniyani, B.~Huberman.
\newblock \emph{Search in power-law networks}.
\newblock Physical review E \textbf{64}(4), 46135 (2001).
\newblock \doi{10.1103/PhysRevE.64.046135}

\bibitem{Bianconi:Barabasi:2001:Competition_and_multiscaling}
G.~Bianconi, A.L. Barab{\'a}si.
\newblock \emph{Competition and multiscaling in evolving networks}.
\newblock Europhysics Letters \textbf{54}(4), 436 (2001).
\newblock \doi{10.1209/epl/i2001-00260-6}

\end{thebibliography}
\end{document}